\documentclass[1p]{elsarticle}
\usepackage{longtable}
\usepackage{pdflscape}
\usepackage{multirow}
\usepackage[english]{babel}
\usepackage[utf8x]{inputenc}
\usepackage{amsmath}
\usepackage{amssymb}
\usepackage[fixamsmath]{mathtools}
\usepackage[inline]{enumitem}
\usepackage{float}
\usepackage{graphicx}  
\usepackage{caption,subcaption}
\usepackage{tabularx}
\usepackage[pdftex,dvipsnames, x11names]{xcolor}  
\usepackage[pdftex,plainpages=false]{hyperref}
\hypersetup{colorlinks=true, breaklinks=true}

  \makeatletter
\def\hlinewd#1{
        \noalign{\ifnum0=`}\fi\hrule \@height #1 \futurelet
           \reserved@a\@xhline}
           \makeatother
           
\begin{document}
\title{Cloud Digital Forensic Readiness: An Open Source Approach to Law Enforcement Request Management}

\author[aa]{Abdellah Akilal\corref{cor1}}
\ead{abdellah.akilal@univ-ubs.fr}
\address[aa]{Université de Bretagne Sud, Vannes, France}
\author[mtk]{M-Tahar Kechadi\corref{cor2}}
\ead{tahar.kechadi@ucd.ie}
\address[mtk]{School of Computer Science and Informatics, University College
 Dublin, Ireland}
\cortext[cor1]{corresponding author}


\begin{abstract}
    Cloud Forensics presents a multi-jurisdictional challenge that may undermines
    the success of digital forensic investigations (DFIs). The growing volumes
    of domiciled and foreign law enforcement (LE) requests, the latency and
    complexity of formal channels for cross-border data access are challenging
    issues. In this paper, we first discuss major Cloud Service Providers (CSPs)
    transparency reports and law enforcement guidelines, then propose an abstract
    architecture for a Cloud Law Enforcement Requests Management System
    (CLERMS). A proof of concept of the proposed solution is developed,
    deployed and validated by two realistic scenarios, in addition to an
    economic estimation of its associated costs. Based on available open source
    components, our solution is for the benefit of both CSPs and Cloud Service
    Consumers (CSCs), and aims to enhance the due Cloud Digital Forensic
    Readiness (CDFR).
\end{abstract}
\maketitle
\section{Introduction}
\label{sec:introduction}
Major cloud service providers (CSPs) are legally domiciled in the USA, but do
have subsidiaries around the world~\citep{googlocation2017, fbinfo2017}. CSPs
consumers, data, and partners are by de facto scattered around the globe. For
example, a Spanish users' data may be stored in a USA datacenter and
processed through an Irish facility. Therefore, data localization is a
challenging issue especially in case of incidents or cybercrimes.

Moreover, law enforcement (LE) access to digital evidence depends entirely on
the CSPs trustworthiness and the complexity of the associated legal procedures
\citep{tcy2015criminal, manral2019systematic}. In fact, CSPs' LE guidelines
point out the need to foreign LE to address their requests through formal
channels such as the Mutual Legal Assistance Treaty (MLAT)
\citep{funk2014mutual}. The assertion of a jurisdiction during an investigation
---\emph{in some instances}--- is problematic. In fact, even if the LE is
domiciled in the same jurisdiction as the CSP, the potential digital evidence
may be located overseas, and require cross-borders data access mechanisms
\citep{svantesson2015access, goog2017digital, mulligan2018cross}.

The multi-jurisdictions issue is one of most persistent Cloud Forensic (CF)
challenges \citep{ruan2011cloud, simou2016survey,manral2019systematic,
herman2020}. Nowadays, CSPs are being tired between domestic and foreign
legislations \citep{goog2017digital}. There is an abundant literature on the
multi-jurisdictional issues
~\citep{walden2013accessing,koops2014cyberspace,svantesson2016law,daskal2018microsoft,Abraha2019}.
However, the existent propositions focus mainly on the legal standpoint, because
it is primarily a legal challenge.

Nonetheless, we inhere, aim to propose a technical and organizational solution
to simplify the LE request handling process. Our main objectives are: (1) To
facilitate the communication between the requester (LE) and the responder, (2)
Ensure transparency in handling a LE request, (3) Enhance the due forensic
readiness capabilities of the responder.

The main contributions of this paper: (1) Analysis of CSPs transparency reports
and law enforcement guidelines, (2) Proposing an abstract architecture for a
Cloud Law Enforcement Request Management System, (3) Design, development and
deployment of an open source based prototype, (4) Validation of the
prototype with two hypothetical scenarios, (5) Economic cost assessment of a real
world deployment.

The rest of this paper is structured as follows: In Section 2 we present some
background on Cloud Forensics, Cloud Digital Forensic Readiness, Multi-jurisdictions,
formal channels for cross-borders digital evidence access, and LE requests
management systems; In Section 3, we state the main problem addressed in this
paper; In Section 4, we clarify the motives and scope of this study.
Section 5 presents a detailed description of the Cloud Law Enforcement
Requests Management System. Section 6 provides a discussion on the opportunities
that may arise from the proposed solution and its limits. Finally, we conclude
this study with some perspectives and future work in Section 7.
\section{Background}
\label{sec:background}
This section gives an overview of the following concepts: Cloud Forensics (CF),
Cloud Digital Forensic Readiness (CDFR), multi-jurisdictions, a description
of formal channels for cross-borders digital evidence access, CSPs transparency
reports and LE guidelines, and some existent law enforcement request management
systems.
\subsection{Cloud Forensics}
\label{subsec:cf}
Cloud Forensics (CF) is the application of digital forensics science to Cloud
computing ~\cite{ruan2011cloud, ruan2013cloud}. Several works have been done
in this field. Researchers' propositions fall in two categories: (a) CF
challenges and solutions classification, (b) contributions that focus on a single
challenge or a category of challenges.

There are numerous CF challenges classifications; \cite{ruan2011cloud} proposed
a CF classification based on three dimensions: Technical, Organizational and
Legal; \cite{pichan2015cloud}
made a survey on the technical CF challenges; \cite{simou2016survey} suggested
a new classification based on the DFI process stages; \cite{herman2020} proposed
the most enumerative classification (65 challenges and 9 primary categories).
And more recently, \cite{manral2019systematic} presented a systematic survey on
CF challenges, proposed solutions, artifacts identification, forensic tools, and
research gaps.

In regards to the second category, some research works focused on challenges,
such as logging, live forensics, and timeline reconstruction. For example,
\cite{zawoad2016trustworthy} presented a forensic enabled architecture built on
the top of the openstack platform \cite{openstack};
\cite{battistoni2016cure} worked on reliable timeline reconstruction;
\cite{Irfan2016framework} presented a framework for Cloud digital evidence
collection and analysis using a Security Information and Event Management (SEIM)
tool; And finally, \cite{zawoad2016towards} proposed a
Secure-Logging-as-a-Service (SecLaaS) solution.
\subsection{Cloud Digital Forensic Readiness}
\label{subsec:cdfr}
Digital forensic readiness (DFR) is the ability to maximize an environment's
potential to use digital evidence, whilst minimizing the costs of an
investigation. Introduced by~\cite{tan2001forensic}, it was then clearly
stated as a ten step process by~\cite{rowlingson2004ten}. Seminal works in this
topic have been done by \cite{endicott2007theoretical, taylor2007specifying,
grobler2007digital, dilijonaite2017digital}. Cloud Digital Forensic
Readiness (CDFR) in counterpart is the application of DFR in Cloud computing
environments. A detailed and concise definition was provided by
\cite{de2013cloud}. Enhancing the CDFR was the subject of many
propositions; \cite{trenwith2013digital} exposed a proof of concept tool for a
centralized Cloud logs; \cite{kebande2015adding} added event reconstruction to a
CDFR model; \cite{kebande2018novel} proposed a Cloud Forensic Readiness as a
Service (CFRaaS) model. Finally, one of the most significant advancements in DFR
and CDFR was the introduction of the \emph{``Forensic-by-design''} paradigm
by \cite{ab2016forensic, ab2017cloud}.
\subsection{Multi-jurisdictions}
\label{subsec:multij}
Cloud computing elasticity induces data localization challenges. In fact, data
may be stored, processed, and mirrored across multiple jurisdictions. As for
data, Cloud services are also consumed by users across the globe.
Multi-jurisdictions in its simplistic form may be sketched as follow:
\emph{How may a Spanish LE agent investigate a cybercrime against Spanish
victims committed by a USA cybercriminal resident in the UK, where the potential
evidence is stored in Ireland, and processed in Asia region data centers that
belong to a USA domiciled CSP?}

Furthermore, the multi-jurisdictions challenge goes beyond the territoriality,
or nationality of customers and providers. In fact, from a legal perspective,
a \emph{``Jurisdiction"} may take at least three forms as stated
by \citealt[pg.3]{svantesson2015access}.

In the case of cross-borders investigations, international law experts specify
two forms of cooperation ---\emph{``formal channels"} and \emph{``informal
channels"}--- between a foreign LE and a CSP
\citep{walden2013accessing,koops2014cyberspace}. Additionally,
\cite{walden2013accessing} formulated four possible courses of action for
foreign LE \citep[pg.55]{walden2013accessing}. Even so, major CSPs are still
restricting foreign LE requests to the formal channel through Mutual Legal
Assistance Treaty (MLAT) or Rogatory letter.
\subsection{Formal channels for cross-borders data access}
\label{subsec:mlat}
There are three formal channels for cross-borders data access: (1) Mutual Legal
Assistance Treaties (MLATs) \citep{funk2014mutual}, (2) Rogatory letters, and
(3) the Clarifying Lawful Overseas Use of Data (CLOUD) act \citep{cloudact2018,
cloudact_ress}. While the two first mechanisms have been there for decades, the
last act was enacted in March 2018.

The MLAT channel is primarily used by LE, and the Rogatory letters are often
used by non-government litigants \citep[see][pg.5, pg.17]{funk2014mutual}. Even
if, the MLAT process is theoretically simple \citep[see][pg.2]{lin2017cross}, in
practice it faces challenges ,such as procedure complexity, processing latency,
guideline issues, and requests processing capacity
\citep{doj2014budget,tcy2014report,james2016survey,woods2017mutual}.

\begin{figure}[!ht]
    \begin{center}
        \includegraphics[width=.95\linewidth]{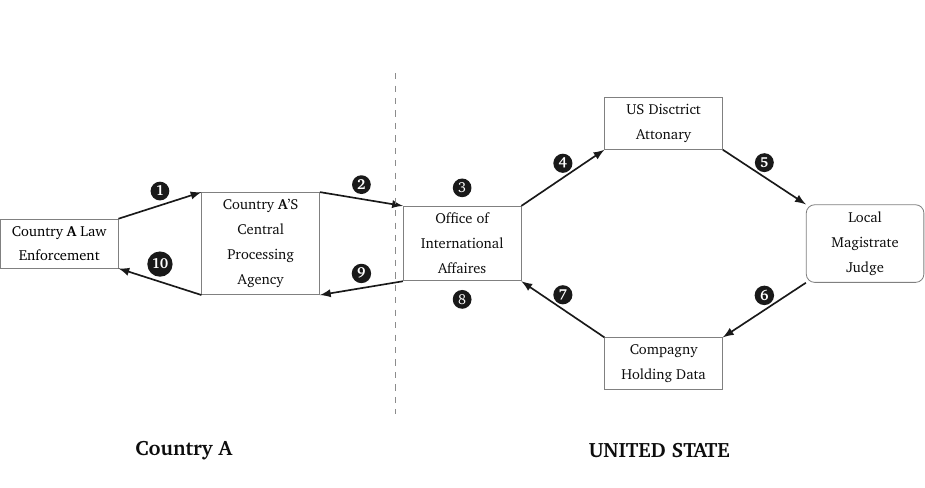}\\
    \end{center}
    \caption{Example of a US MLAT process for Electronic Evidence. (adapted from
    \citep{lin2017cross}}\label{fig:mlat_sch}
\end{figure}

As described in \cite{lin2017cross} (as shown in Figure.
\ref{fig:mlat_sch}) ,
when a country A law enforcement needs access to cross-borders data, the request
is sent to its associated central processing agency in step 1. Afterwards, the
central processing agency sends the request to the Office of International
Affairs (OIA) at the US department of justice (2). The OIA reviews the foreign
request in phase 3 and works with the U.S district Attorney's office (4). The
request is then sent to a local magistrate judge for review (5). If the judge
deems the request receivable, a warrant is served to the company which holds the
request data (6). Complying to the warrant, the company provides the requested
data to the OIA (7) for post processing overview (8). Finally, the requested
data is sent to the country's A central processing agency which forwards it to
the law enforcement in the last step (10). The whole process takes from six
weeks to ten months.

Finally, the CLOUD act aims to address some of  the MLATs inefficiencies to
speed up the access to digital evidence \cite[see][page
333]{bilgic2018something}. It authorises bilateral agreement between the US
and a \emph{trusted} foreign partner to obtain direct access to digital
evidence, wherever they are located. However, eligible foreign countries must meet some
requirements such as the protection of privacy and civil liberties during the
data-collection activities. In fact, the \emph{``trusted foreign partner''}
condition has some political ramifications. As for now, only some countries such
as the U.K have gained benefit from this mechanism. Moreover, there are several
research works discussing the impact of this new legislation on the existent
formal channels, privacy, and international laws
\citep{mulligan2018cross,daskal2018microsoft,Abraha2019}.
\subsection{Transparency reports and LE guidelines}
\label{subsec:transparency}
A law enforcement request may vary from disclosure, preservation, removal, and
testimony. On counterpart, CSPs may approve, reject or even challenge a request
\citep{svantesson2015access,eff2015mswarrant}. CSP's response to received LE
requests is mainly based on: (a) domestic jurisdiction laws, (b) regulations,
(c) internal policies \citep{goog2017leguid, fb2017leguid}, (d) associated legal
procedure, e.g., \emph{Subpoenas, Court orders, Search warrants, etc.}
\citep{yah2017leguid}, and (f) the request status (emergency or not). In fact,
emergency requests  are handled promptly, and processed with the due priority
\citep{amzn2014leguid,fb2017leguid}.

Approved LE disclosure requests may lead to the communication of either
\emph{``Non Content Data''} or \emph{``Content Data''}, if available data is
found for the request stated period
\citep{goog2017leguid,fb2017leguid,amzn2014leguid}.
Statistics on disclosed users' information are published in CSPs transparency
reports, which are often heterogeneous. In fact, only Microsoft
\citep{mstrreport2017} and Yahoo \citep{yah2017leguid} seem to use a similar
format. Additionally, CSPs also publish guidelines for LE
\citep{yah2017leguid,ms2017leguid,goog2017leguid,fb2017leguid,amzn2014leguid}.
Table. \ref{tab:trans_leguid_attr} provides a summary of transparency and law
enforcement guidelines attributes.
\begin{table}[H]
\scriptsize{
\begin{center}
    \begin{tabularx}{\linewidth}{ l  |  X }
        \hlinewd{1pt}
        \multicolumn{2}{c}{\textbf{Transparency reports}}\\
        \hlinewd{1pt}
        \textbf{Attribute} &\textbf{Description}\\
        \hlinewd{1pt}
        Volume& Number (percentage) of received requests, Users' impacted
        accounts, and delivered responses (accepted, rejected). In case of
        accepted  requests statistics on \emph{``content data''} vs \emph{``no
        content data''} responses are also provided.\\
        Requester& Type of requester authority: Law Enforcement Agency (LEA),
        FISA\\ 
        Legal procedure& LE requests associated documents (subpoena, court
            order, search warrant, etc.).\\
        Localization& Requester geographical origins, either by country or a
            comparison between domestic country vs. Foreign.\\
        Status&Specification on requests regime (emergency or not).\\
        Objective&Requested actions on target's data (disclosure, removal,
            preservation and testimony).\\
        Historic&Indications on past transparency reports.\\
        \hlinewd{1pt}
         \multicolumn{2}{c}{\textbf{Law Enforcement Guidelines}}\\
         \hlinewd{1pt}
        Legal requirements& Request legal ground of acceptance.\\
        Emergency exception& Exceptions to the legal requirements for emergency requests.\\
        Suited actions& Type of receivable requests, and suited actions on
        data. In the case of a preservation request, the associated preservation
        delay.\\
        Target&Targeted user required identity (Id, Account, Email, etc.).\\
        Target notification&Target notification policy, when and how a CSP may
        (or not) notify a user about an issued government request on its data.\\
        Requester identity&Information required from a requester to show
        authenticity and authentication.\\
        Submission means&Permissible means for request submission (email, mail,
        fax or online).\\
        Costs reimbursement&Regime, and exception on cost reimbursement to due
        business disturbances and request compliance induced fees.\\
        \hlinewd{1pt}
    \end{tabularx}
    \caption{CSP's Transparency reports and law enforcement guidelines attributes}
    \label{tab:trans_leguid_attr}
\end{center}
}
\end{table}
\subsection{LE requests management system}
\label{subsec:lems}
To the best of our knowledge, there is no previous work on law enforcement
requests management systems. Certainly, major CSPs do have their own solutions to
manage and process LE requests. However, LE pre-submission request is  only
permitted via Fax or Email. As far as we know, only Facebook and Google
\citep{fbportal2020, lers2020} propose an online requests pre-submission portal
for the exclusive usage of LE agents.

In the following section, we provide details on some of the law enforcement
request handling challenges.
\subsection{Legal request processing Issues}
\label{subsec:le_req_chall}
As stated in Subsection. \ref{subsec:multij} multi-jurisdiction is a persistent
challenge in Cloud Forensics and more specially in cross-border data access.
However, legal request processing faces also other challenges that are amplified
by legal consideration, such as the following:
\begin{enumerate}
    \item \textbf{Volumes}: There is for sure a significant growth in LE
        requests. For example, the number of legal requests for users' information
        disclosure submitted to Google grew from $165,894$ in 2019 to
        $217,424$ in 2020. If this is the current state for CSPs, the situation
        may be worse considering the central entity---Office of International
        Affairs-- that evaluates foreign LE requests through ---via Formal
        channel (MLAT)---. Indeed, in the United State, the OIA state that:
        \begin{quote}
            Over the past decade the number of requests for assistance from
            foreign authorities handled by the Criminal Division's Office of
            International Affairs (OIA) has increased nearly 60 percent, and the
            number of requests for computer records has increased ten-fold.
            \citep[see page 1]{doj2014budget}.
        \end{quote}
        The growing number of LE requests implies a need for automation for both
        the responder and the central entity that evaluate the foreign request.
    \item \textbf{Delays}: Latency in processing LE requests is observed in two
        cases: (1) At the CSPs level, in fact a request is not processed until
        the reception of the associated legal documents via mail, (2) in case of
        foreign request via formal channel where a nine step process is enacted
        and which involve many instance other then the CSP (see Subsection. 
        \ref{subsec:mlat}). The MLAT process takes from six weeks to ten months
        on average \citep[see page 3]{lin2017cross}. 
    \item \textbf{Transparency}: The lack of transparency is observed at least
        in two cases: (1) Generally, CSPs do not provide any technical
        capabilities for LE to monitor the process flow of their submitted
        requests ---with the exception of those that propose an online
        portal---, additionally, CSPs do not provide response to testimony
        requests, (2) Considering foreign request, formal channel procedure are
        so complex and involve several parties that it is difficult to monitor
        the handling of the submit request. Furthermore, in case of requests
        leading to the collection of digital evidence, the admissibility of
        those collected evidence stands only on the trustworthiness of the CSPs, and
        may be challenged in a foreign court of law.
        Indeed, some researcher have already pointed out this possibility:
        \begin{quote}
            It is important to recognize that data obtained from a cloud-based
            service may be excluded from use in court proceedings on a number of
            grounds \citep[see][pg. 64]{walden2013accessing}.
        \end{quote}
    \item \textbf{Procedures}: Law enforcement guidelines are mainly in English
        or translated into a few languages. Assisting LE in the submission process
        is crucial. However, CSPs do not provide detailed procedure for request
        submission, nor do they provide standard forms. Assuming that LE
        agents are fully trained or do have the necessary training to make such
        a request may lead to submission being rejected due to error or
        misinformation. As for the foreign requests, procedures are more
        complex. Foreign LE training on how to formulate request through formal
        channel is required as stated by \citep{doj2021audit}:
        \begin{quote}
            Coordinate with OIA to develop a plan to improve its training and
            outreach efforts including
            considering the creation of an external site of resources for
            foreign authorities.\citep[see page 29]{doj2021audit}
        \end{quote}
\end{enumerate}

A stated above, there is certainly a dependency between the legal and technical
aspects of LE requests processing. The complexity of legal procedures for
cross-border data access induce some technical challenges, such as latency,
volume and transparency. On the other hand the lack of automation in both
responder and central entity (OIA) request processing induce an increase of the
cases backlog and therefore latency in legal reviews of the formulated request
both domestic and foreign.

The need for a scalable technical solution for LE requests processing for both
responder and central entity ---in case of a foreign request-- is persistent.
Indeed, even the OIA have expressed the need for a reform in MLAT processing and
the adoption of scalable case management system (CMS) as expressed below in
\citep{doj2021audit}:
\begin{quote}
    Coordinate with CRM ITM to ensure OIA has access to CRM's Oracle Apex
    platform and support the automation of OIA's team trackers and leadership
    dashboards. \citep[see page 29]{doj2021audit}.
\end{quote}

In the following section, we describe clearly the main problem that we aim to
resolve in this study.
\section{Problem statement}
\label{sec:problem_statement}
Certainly digital evidence access in Cloud computing environments is very
challenging (see Subsections. \ref{subsec:multij} and \ref{subsec:mlat}).
Technical issues are amplified by legal aspects. If the handling of domestic LE
requests seems straightforward, foreign requests in counterpart are subject
to more complex procedures. Among the aforementioned challenges, trust and
transparency seem to be the most difficult to resolve. Moreover, these two
aspects (transparency and trust) are also required to ensure the soundness of
the conducted DFI processes and the admissibility of the collected digital
evidence.

After the analysis of transparency reports and LE guidelines (see
Subsection. \ref{subsec:transparency}), it appears that there is a need to
either maintain the current state of LE request handling ---processes and
associated technical capabilities--- waiting for advancement in legal frameworks
establishment especially in case of foreign LE requests, or anticipate the venue
of these
\emph{``legal frameworks''} by designing and developing new technical
capabilities that may resolve some challenges and attenuate the complexity of
other ones. In fact, some major CSPs are already providing an online portal for LE
request pre-submission (see Subsection. \ref{subsec:lems}).

If CSPs possess the resources needed to the design and the development of
associated LE requests processing technical capabilities, other organisations
(eg.,  Small and Medium Sized Enterprise) that are outsourcing part of their
information system to a CSP may not. And, in some circumstances those
organisations may be asked to respond to legal requests.

So, in abstraction of the nature of the responder (CSP or not), in the rest of
this  paper we aim to answer the following question:

\emph{What are the required technical capabilities that a responder should
possess in order to handle transparently legal requests in accordance with
specified guidelines, laws, and regulations ?}

\section{Motives and Scope}
\label{sec:motives_and_scope}
To the best of our knowledge, there is no prior work addressing specifically cloud legal
request  management system.  Moreover, the  only existing  systems are  proprietary (CSP
in-house made), which  are exclusively used by the law  enforcement agencies without any
public or published documentation.

Therefore, our main motivation in this research is to establish an affordable, scalable,
and open source based  technical solution that may ease some of  the LE request handling
challenges. More  precisely, we  are aiming  towards the design  and the  development of
solutions that fit any organisation ranging  from public CSPs to the smallest enterprise
which outsources its Information System (IS) from  a CSP. The main objectives to achieve
are:

\begin{enumerate}
    \item Facilitate the communication between the requester and the responder
        and a ensure a transparent request handling,
    \item Provide the responder with a minimal DFR baseline to guarantee a sound
        DFI process and the admissibility of the collected digital evidence.
\end{enumerate}

\section{The Proposed solution}
\label{sec:clerm_framework}
Upon statement of the main problem (see Section. \ref{sec:problem_statement}),
in this section, we propose a technical solution for some of the LE requests
handling challenges. Therefore, we will address the following points:
\begin{enumerate}
    \item Analysis of LE request handling workflow.
    \item Proposition of abstract architecture.
    \item Analysis of the system requirement.
    \item Design, development and deployment of a prototype.
    \item Validation of the proposed system with two hypothetical
        scenarios.
    \item Economic assessment of the associated deployment costs.
\end{enumerate}
\subsection{LE request processing flow}
\label{subsec:le_request_proc_flow}
CSPs governance is based on: (1) business term of services, (2) applicable
domestic jurisdiction laws, and (3) compliance to regulations and standards.

That said, governance is also asserted through organizational policies and
technical capabilities. Therefore, the specification of the internal workflow,
data flow, and external partners' communication channels are vital. In this
context, existent incidents response and digital forensic investigation policies
may be enhanced with a new law enforcement policy that specifies the
communication with LEs. Furthermore, on the technical side, the implementation
of a law enforcement request management system will ease the
CSP-LE communication and ensure a transparent requests processing.

\begin{figure}[H]
    \begin{center}
        \includegraphics[width=.8\linewidth]{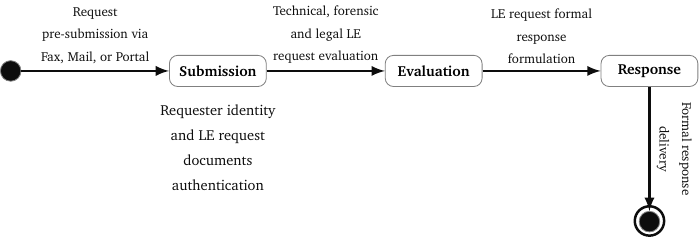}\\
    \end{center}
    \caption{LE request processing workflow}\label{fig:workf}
\end{figure}

Based on available CSP LE guidelines \citep{goog2017leguid, fb2017leguid}, a
typical LE request processing workflow (Figure. \ref{fig:workf}) comprises 3
major steps: submission, evaluation and response.

Even if there are some considerations in the processing of some LE requests (see.
Subsection. \ref{subsec:transparency}) (e.g., an emergency request is considered
as an exception to the LE guideline and handled promptly), the above cited
stages are still present and the handling particularities appear in some
procedures and tasks.
In the following details on the LE request handling stages:
\begin{enumerate}
    \item \textbf{Submission}: we observe that major CSPs accept only requests
        submitted via Fax or Email. Even if there are some CSPs that offer an online
        portal for request pre-submission, the effective request evaluation is
        only initiated at the reception of the legal documents via the mail.
        However, in case of emergency or preservation requests, some actions may
        be initiated (anticipated) in due respect to the adopted law enforcement
        policy.

    \item \textbf{Evaluation}: upon effective reception of the request's legal
        documents, a team composed of law assistance, IT manager, incident
        response members, and forensic experts may gather to evaluate a possible
        response based on pre-established internal law enforcement policy. In
        case of approved disclosure (or preservation) requests an escalation to
        a full DFI is required. In such a case, collected digital evidence, chain
        of custody documents and other DFI reports may be included within the
        formal response to the requester.

    \item \textbf{Response}: a formal answer is transmitted to the requester
        (via email or mail). Approved requests may lead to: (1) the application
        of the requested actions (preservation, disclosure, deletion) on the
        target's data, (2) notification the target (in some cases),
        establishment of an invoice for the costs reimbursement. Moreover, a
        responder may also reject a request or even challenge it (see
        Subsection. \ref{subsec:transparency}). Finally, we observe that
        major CSPs do not respond to testimony requests and offer only a
        certificate in exchange of expert testimony.
\end{enumerate}
\subsection{Architecture definition}
\label{subsec:architecture}
\begin{figure}[!ht]
    \begin{center}
        \includegraphics[width=.99\linewidth]{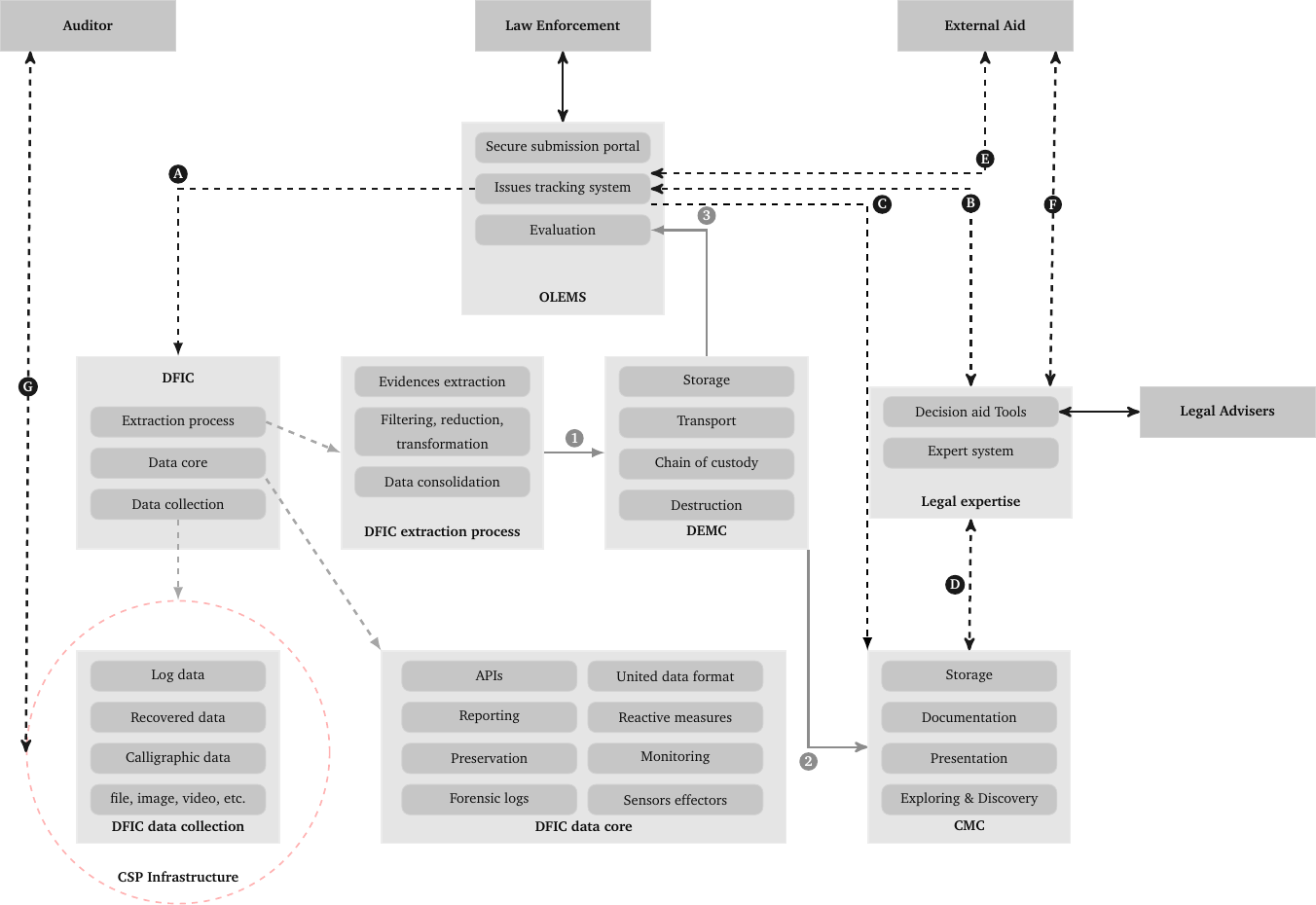}\\
    \end{center}
  \caption{An abstract architecture for a law enforcement requests
        management system. OLEMS: Online Law Enforcement Management System.
        DFIC: Digital Forensic Investigation Capabilities. DEMC: Digital
        Evidence Management Capabilities. CMC: Cases Management
        Capabilities.}\label{fig:arch}
\end{figure}

With the established LE request processing workflow (Figure. \ref{fig:workf}), the
information gathered from CSPs transparency reports, LE guidelines analysis
(Section. \ref{subsec:transparency}), and the composition of the evaluation
team, we may sketch an abstract architecture for a law enforcement requests
management system (see Figure. \ref{fig:arch}).

After the submission phase, if an escalation to a full DFI is required, the
Digital Forensic Investigation Capabilities (DFIC) are activated through the
connection (A), and a new case is opened in the Case Management Capability (CMC).
Legal expertise may also be solicited through the connection (B). Once the DFIC
is activated, a sound DFI process is engaged. First, the collection of potential
digital evidence is made across the CSP's infrastructure using the DFIC
\emph{data core} sub-capabilities, then evidence are extracted with the DFIC
\emph{extraction process}, and securely managed by the Digital Evidence
Management Capabilities (DEMC).

When the required digital evidence are available, they are included in the
associated case via the connection (2), and the findings transmitted to the OLEMS
for reporting via the connection (3). Legal advisor's having a secure access to
the CMC via the connection (D) may establish their reports based on elements of
the current case, and the knowledge that they may have discovered from similar past
cases.
Once the technical, forensic, and legal reports are available, the CSP
formulates a response to the LE, either via the OLEMS or any available legal means.

The following sections provide details on the different modules of the proposed
architecture.
\subsubsection{Online law enforcement management system (OLEMS)}
\label{subsubsec:olems}
This group of capabilities aims to improve the interactions between: (1) CSP and
LEs, (2) involved teams in a request processing, and it contains the following:
\begin{enumerate}
    \item A secure pre-submission portal to accelerate in some instances the
        evaluation and response delays. In fact, even if the authenticated
        documents are required, the emergency and preservation requests may
        benefit from proactive measures dictated by the evaluation team;
    \item An issue tracking system for the internal staff and the LE to
        monitor the request processing workflow;
    \item Productivity tools (information exchange, virtual meeting,
        redactions tools, etc.) to enhance the collaboration of the involved
        teams (internal/external).
    \end{enumerate}
\subsubsection{Digital forensic investigation capabilities (DFIC)}
\label{subsubsec:dfic}
This group contains the required capabilities that ensure a proper and sound
DFI. Inspired from \cite{de2013cloud} abstract reference architect, it includes
the following:

\begin{enumerate}
    \item Data collection: represents the inventory of the \emph{potential}
        digital evidence sources and the suited collection tools. Therefore, log
        data, available and recovered artifacts are gathered among the several
        layers (physical and virtual) of a CSP' infrastructure from the hardware
        to the application levels.
    \item Data core: regroups the locally and remotely accessible core primitives
        that are required for evidence collection and analysis, monitoring, data
        format unification, log preservation, etc.
\item Extraction process: contains the suited tools for data consolidation,
    filtering, reduction and transformation.
\end{enumerate}
\subsubsection{Digital evidences management capabilities (DEMC)}
\label{subsubsec:demc}
This group of capabilities aims to ensure evidence safeguard and admissibility,
and the maintenance of the chain of custody. As shown in Figure.
\ref{fig:arch}, this group contains four capabilities: storage, transport,
chain of custody and destruction. As for the storage, several formats of digital
evidence exist. These include Raw, DEB and AFF4 \citep{cohen2010,Prayudi2015}. In
addition, there are also some other attributes related to a
storage capability, such as storage area, infrastructure security, and scalability
\citep{Cruz2015}. Concerning the transport, in case of a cooperation among
several forensic experts, a secure transport capability that ensures the
security of digital evidence and the maintenance of the chain of custody is
required. Indeed, the maintenance of the chain of custody is vital for the
admissibility of the collected evidence. In this scope contribution such as the one provided by \cite{Chew2024} may be considered. Finally, depending on the associated
jurisdiction, the destruction of the collected evidence may be considered.
\subsubsection{Cases management capabilities (CMC)}
\label{subsubsec:cmc}
Usually, when a request is processed, it generates a case that contains elements,
such as a digitized version of the request papers, evaluation reports,
briefings notes, artifacts, evidence, and evidence custody documents, etc. CMC
capabilities aim to ensure the storage and the preservation of the
aforementioned elements.
In the following section, we provide details on the proposed system requirements.
\subsection{System requirements}
\label{subsec:implementation}
This section aims to implement a Proof of Concept (PoC) of the proposed
architecture (Section. \ref{subsec:architecture}). Our methodology is project
driven. However, to optimize the costs of the implementation, we adopted the
following guidelines: (1) candidate solutions must be open source, or
responder's in-house made, (2) scalability, (3) portability and
interoperability. On the other hand, note that there some other non-functional
system requirements, such as (4) Cloud Digital Forensic Readiness
(DFR) (\emph{i.e., ability to collect admissible digital evidence while optimising
the costs of an investigation}), (5) ensure a transparent LE request handling,
and (6) Trust.

Even if the chosen modules are open source, there are still costs related to
the Cloud deployment. Nonetheless, in the following sections, we first assess the
requirements for each sub-module (Figure. \ref{fig:arch}) in Subsection.
\ref{subsec:proto_dev} and provide an economic assessment of the proposed
solutions costs in case of a real world enterprise deployment in Subsection.
\ref{subsec:economics}. 
\subsection{Prototype design \& development}
\label{subsec:proto_dev}
\begin{figure}[H]
    \begin{center}
        \includegraphics[width=.95\linewidth]{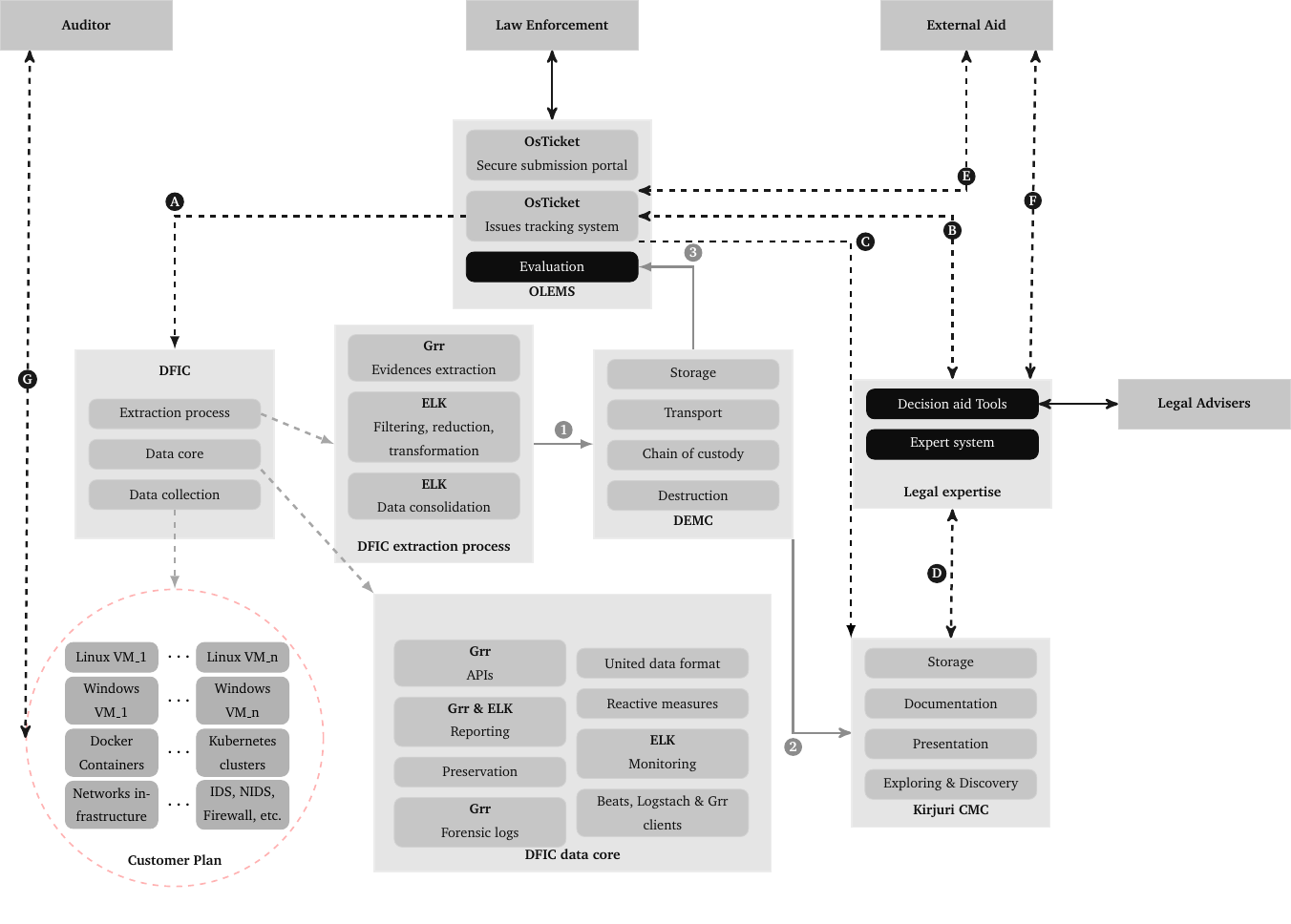}\\
    \end{center}
    \caption{A Cloud Law Enforcement Request Management System (CLERMS)
    implementation using open source components. The connection A, B, C, 1,2 and 3
    represent the interactions between sub-modules as specified in
    in the abstract architecture (Section.
    \ref{subsec:architecture}). Black colored box are capabilities that are not
    implemented in the prototype.}\label{fig:imp}
\end{figure}

Starting from the OLEMS, a secure
LE requests pre-submission portal is required. The request processing may be
tracked through a tickets system or any help desk solution. Therefore, there is
a need for Email and SMS notifications; Tickets assignment and management;
Dashboards and reporting features. There are many open source solutions that
fulfil the cited requirements, such as \cite{osticket} and \cite{zammad}. In
this study, we opted for the osTicket solution as sketched in Figure. \ref{fig:imp}.

At the reception of the request papers, they are scanned and stored for
legal and technical evaluation. So, there is a need for electronic documents
management and knowledge discovery solutions.

If a request is approved,  an escalation to a full DFI is initiated. Therefore, the
usage of dedicated tools for live forensic, evidence and log data collection
(acquisition), analysis and examination are vital. These tools must support multiple
types of artefacts, and ensure secure evidence storage.

For the monitoring and log analysis, we opted for the Elasticsearch Logstash
Kibana stack \cite{elastic}. Events, network packets
and logs are collected via agents (Beats) deployed on the customers plan and
forwarded to the Elasticsearch cluster in real-time. The Kibana dashboards help
in monitoring, gaining insight, and hunting threats. For the incident response
and live forensic, we opted for the Google Rapid Response (Grr) solution
\citep{cohen2011distributed, moser2013hunting, cruz2015scalable}.

Finally, for case management, we opted for the open source solution
\cite{kirjuri} that provides support for (1) investigators and evidence
management, (2) investigation and chain of custody documents (request documents,
evaluation reports, evidence analysis reports, etc.) management, digital
evidence storage. However, digital evidence storage and management may be
feasible through a separated data store, such a MangoDB cluster as suggested in
\citep{cruz2015scalable}.

As for the \emph{evaluation} module and the legal capabilities they are not
considered in the developed prototype as they may be ensured by offline
(in-house made or open source) solutions. Mainly consisting of an
assistance and aid tool for reports editing (in case of the evaluation module),
and a knowledge database and electronic document discovery for the legal
capabilities, these elements may be separated from the connected system.

The following section provides details on the deployment of a proof of concept
prototype of the proposed solution on an IaaS infrastructure.
\subsection{Cloud deployment}
\label{subsec:deployement}
\begin{figure}[H]
    \begin{center}  
      \includegraphics[width=.9\linewidth]{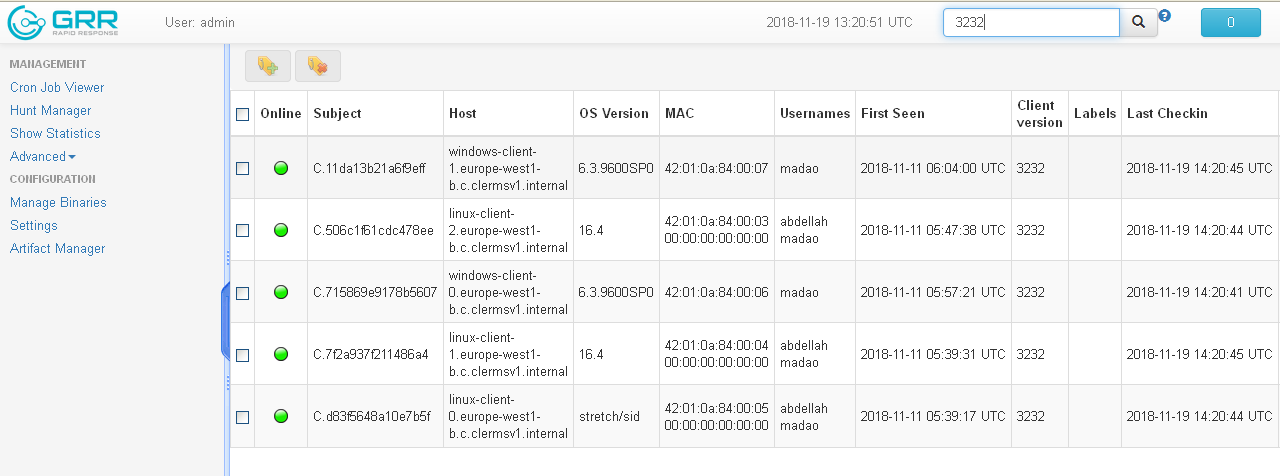}
    \end{center}
  \caption{Grr-server administration user interface.}
  \label{fig:grr}
\end{figure}

\begin{figure}[H]
    \begin{center}
        \includegraphics[width=.9\linewidth]{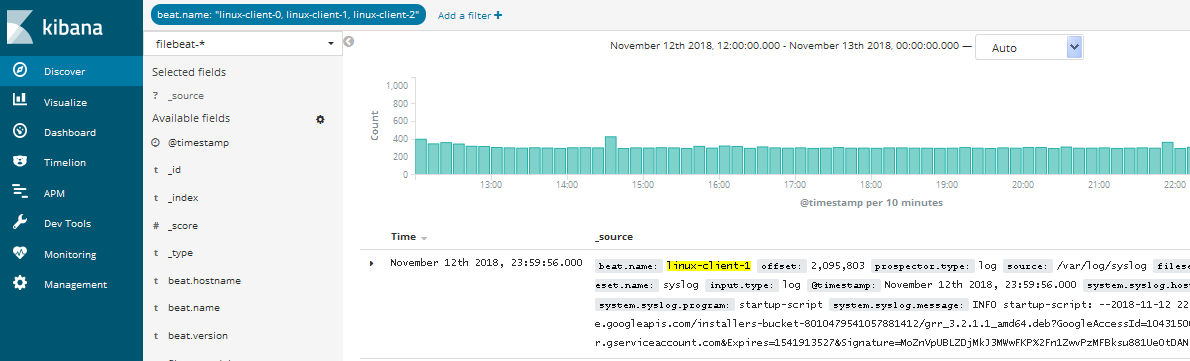}
    \end{center}
    \caption{Monitoring, logs ingestion, and threat hunting via Kibana
    dashboards.}
  \label{fig:kibana}
\end{figure}

For the deployment of our prototype, we opted for a Cloud (IaaS) platform
\cite{GoogleCloud} as shown in Figure.
\ref{fig:imp}. The allocated resources for some modules, such as the Grr-server,
Elasticsearch, and Osticket, depend on the costumer plan (i.e, number of
targeted nodes). In fact, depending on the demand ---growth or decrease in the
number of client nodes--- the deployment of these modules may be scaled up or
down.
\begin{figure}[H]
    \begin{center}
        \includegraphics[width=.9\linewidth]{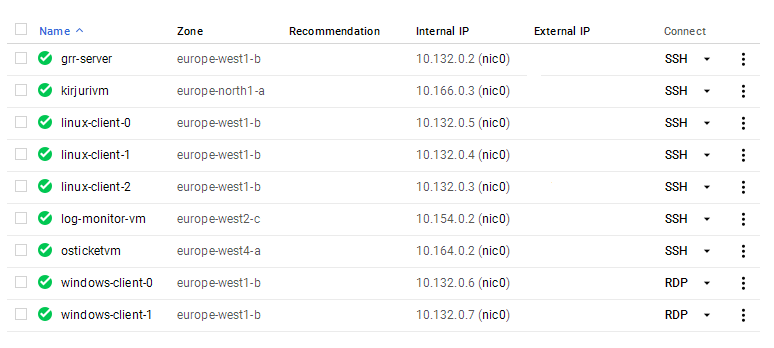}
    \end{center}
    \caption{CLERMS deployment on an IaaS infrastructure.}
  \label{fig:deploy}
\end{figure}
In our case, the customer plan comprises 5 Virtual Machines (VMs) (3 were Linux
based, and two Microsoft Server 2010 instances). For larger deployment, the Grr
documents \cite{grrdoc} specify the type and requirements for the Grr-server
deployment. So, in our case a single VM instance was sufficient (see Figure.
\ref{fig:deploy}); The Grr-server administration interface is shown in Figure.
\ref{fig:grr}.
Logging and monitoring are done via agents (LogStach agents, Beats) on the
customer's VMs plan. The collected information is forwarded to the ELK cluster.
In the production mode ---real world case usage--- the ELK cluster deployment
requires at least three nodes. However, in our case a single node was
sufficient. The Figure. \ref{fig:kibana} shows log visualization through Kibana
dashboards.

\begin{figure}[H]
    \begin{center}
     \includegraphics[width=.8\linewidth]{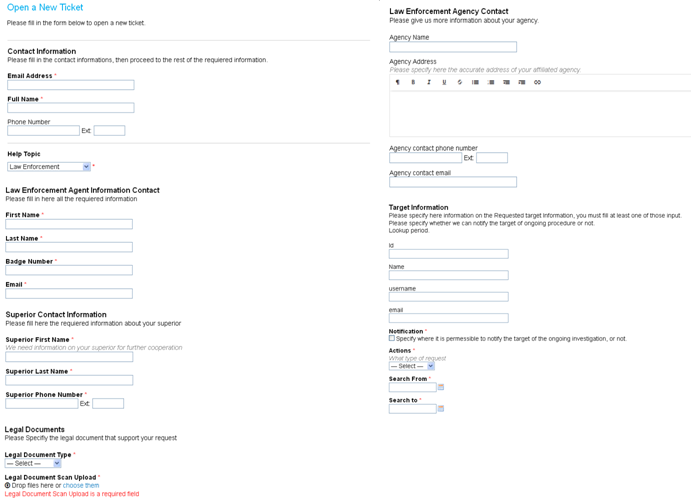}
    \end{center}
    \caption{Online LE request pre-submission portal}
  \label{fig:ticket}
\end{figure}

The online request pre-submission is feasible through an osTicket instance (Fig
\ref{fig:ticket}), which was deployed in a single VM instance. In the case of
traffic increase, the scaling up may be achieved through redundancy and load
balancing.

The required information for request pre-submission based on LE guidelines
(see Section. \ref{subsec:transparency}) are: (a) agent contact information, (b)
agent superior contact information, (c) agency contact information, (d) scan of
legal documents that support the request, and (e) target information.

When a request is correctly submitted (see Figure. \ref{fig:ticket}), a
notification is sent to the crisis manager. Consultation is then made with the
legal assistance, and in some cases an investigation is initiated. In case of
escalation, tickets and tasks are forwarded to those concerned by the
investigation. At the conclusion of the investigation, elements related to the
case are forwarded to the case management solution.

\begin{figure}[H]
    \begin{center}
         \includegraphics[width=.8\linewidth]{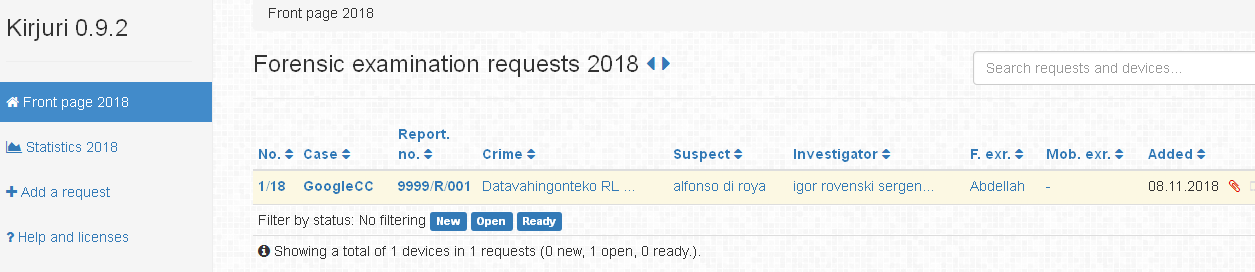}
    \end{center}
  \caption{Kirjuri open source case management User Interface.}
  \label{fig:kirjuri}
\end{figure}

At the reception of the request documents, a new case is opened in the Kirjuri
solution (see Figure. \ref{fig:kirjuri}). The requester and target information,
digital evidence, log, briefing, forensic analysis and examination reports are
securely stored.

The following section provides two hypothetical cases to validate our
proposition.
\subsection{Prototype Validation}
\label{subsec:validation}
This section provides two hypothetical usage scenarios for the CLERMS:(1) user
information disclosure request, (2) content removal request.
\subsubsection{User information disclosure request}
\label{subsubsec:scen1}
A LE agent \emph{``Mike Davies''} is filling a user content information
disclosure request for the target surname \emph{``John smith''}. The target is
suspected of posting illicit content on a discussion forum hosted at
\emph{http://wwww.mydomain.com/fluxbb}. At the request approval, the  crisis
manager decide to escalate the event to a full DFI. Based on the target
associated IP, the investigation team interrogate the customer plan (via the
Grr-servers) for more indications. Afterwards, the forensic expert
may opt for:
\begin{enumerate}
    \item Remote raw disk image acquisition, for further examining and analysis,
        or
    \item Database files acquisition  via the \emph{filefinder} flow (see
        Grr-server administration panel) on the path
        \emph{``/var/lib/mysql/fluxbb''}.
\end{enumerate}

Moreover, when the forensic expert gains the target \emph{``Jhon smith''}
associated registration IP; A comparison with the provided logs from the ELK
stack may confirm whether the illicit content was posted from the target IP
address or not.
Progressively the associated request case will then comprise elements, such
as the digital forensic evidence, logs analysis reports, briefing memos, tasks
assignment, a clear investigation workflow and a proper chain of custody
documentation. Finally, a response to the request may be provided via the ticket
system or by any official means.
\subsubsection{User content removal request}
\label{subsubsec:scen2}
The same LE agent is now suspecting that one of the machines in the customer plan
is hosting a Command and Control program that is in contact with disseminated
malware agents. A removal request with a target machine IP address is then
formulated.
Upon reception of the request, the crisis manager's main objective is the
disinfection of the potential infected machines. A detailed process for such
action may be found in \cite{cohen2011distributed}. However, the forensic expert
may also envision the following: (1) analysis of the processes list flow search
(see Figure. \ref{fig:grr}) for a potential Command and Control process, (2)
proceed to a forensic memory analysis of the suspected machines, and (3) analyze
the logs, SSH transactions, and bash command history, etc. The aforementioned
actions may then help to find the identity of the culprit.

The following section provides an economic assessment of the CLERMS deployment
costs.
\subsection{Economic assessment}
\label{subsec:economics}
The economic optimality of our solution stands on two facts: (1) open source
components, therefore there are no related license fees, (2) Cloud based
deployment for a pay per usage model and scalability (up or down) on demand.
Thus, the CLERMS solution could fit any enterprise or CSPs. Moreover, in
the case of compliance to a LE request, the CSP (requested company) may
formulate a concise, detailed and transparent costs reimbursement invoice based
on the deployment costs (hourly/minutes), man power charges, consulting and
support fees, etc. 

Concerning the deployment related charges, Table. \ref{tab:economic}
provides an assessment of the CLERMS  monthly costs for a customer plan that
consists of 7000 (7K) nodes which is about the size of the infrastructure of a large
enterprise. The costs listed in Table. \ref{tab:economic} are related to
the functioning of the incident response and forensic plan (see Figure.
\ref{fig:imp}), to be accurate. Considered as the baseline configuration for
both the Grrr and ELK deployment in such a case (a customer plan with 7000 nodes),
it may be scaled up and down depending on the usage. Moreover, the responder may
opt for additional digital evidence storage capacity, such as a Mongodb cluster
(more than three nodes with additional storage disks) as specified in
\citep{Cruz2015}.

\begin{table}[H]
\caption{\itshape Economic assessment of  CLERMS  deployment solution for a
    customer plane of 7000 nodes.}\label{tab:economic}
\center\scriptsize
\renewcommand{\arraystretch}{1.3}
\begin{tabular}{rll}
    \textbf{Module} & \textbf{Deployment} & \textbf{Monthly}\\ 
    & requirements & costs (\$)\\ 
    \hlinewd{.5pt}
    Osticket & 1 VM (n1-standard-1)&  24,27 \\
    &(1 vCPU, 3,75 GB Memory)& \\

    Kirjuri  & 1 VM (n1-standard-1)&  24,27 \\
    &(1 vCPU, 3,75 GB Memory)&\\

    ELK & 3 VMs (n1-standard-2)&\\
    &(2 vCPU, 7.5 GB Memory),&165,63 \\
    &500 GB Boot disk,&\\
    &3 persistent disks of 500 GB each
    \cite{esdeploy}
    &\\ 
    Grr-servers & Recommended
    \cite{grrdoc} 5 AWS
    c5a.xlarge (\$0.077 per hour)&1,940.4\\
    &, and one r3.4xlarge
    \cite{awsprice}
    (\$1.328 per Hour)& 6,693.12\\ 
    \hlinewd{.5pt} 
    & \textbf{Total}& 8.847,69\\
\end{tabular}
\end{table}
\section{Opportunities \& limits}
\label{sec:opp_limit}
The proposed solution aims to: (1) ensure a transparent LE request handling, and
(2) enable a responder with a due digital forensic readiness capability in order
to guarantee the soundness of conducted digital forensic investigation and the
admissibility of the collected evidence. Based on open source components it
permits : (a) a reduction of inherent in-house or solution acquisition fees, and
(b) an effortless integration process. Designed to fit any sized responder  (i.e.,
going for a public CSP to the smallest enterprise that is outsourcing  parts of
its own IS to a CSP) the proposed solution will aid a responder to formulate a
concise and precise estimation of LE request processing costs reimbursement
fees.

By ensuring the due digital forensic readiness the proposed solution empower a
responder to conduct sound investigation both in case of a response to a legal
request, and an internal investigation in case of a security incident.  

In regards to legal request processing, the proposed solution induces automation
---and scalability-- that for sure will reduce: (1) the request's volume and
lower its growth rate, (2) latency and expected response delays.

Distributed and scalable technical solutions are not only envisioned for a
responder. Indeed, it is also recommended for the central entity (OIA) that
handles incoming foreign requests via formal channels such as MLAT. Such technical
improvement will for sure reduce the case backlog and processing delays and
therefore accelerate the handling of foreign legal requests. In fact, in an effort to
reform the MLAT processing the US Department of Justices has already expressed
the following recommendation:
\begin{quote}
    Coordinate with CRM ITM to ensure OIA has access to CRM's Oracle Apex
    platform and support the
    automation of OIA's team trackers and leadership dashboards. \citep[page
    29]{doj2021audit}.
\end{quote}

Even if there are promising opportunities that may emerge from the adoption of
the proposed solution there are still some remaining challenges, such as
multi-tenancy and digital evidence collection in PaaS or SaaS service models.
\section{Conclusion and future works}
\label{sec:conclusion}
In this paper, we proposed a Cloud Law Enforcement Requests Management System.
Interests in this proposition come from the growth of law enforcement request
volume, the latency, and the complexity of requests processing.

Our primary goal was to provide CSPs and organizations with an affordable
solution to comply with the received requests. Therefore, we first analysed CSP's
transparency reports and law enforcement guidelines. Second, we sketched a
request processing work flow, enumerated the involved teams, and provided an
abstract architecture. Afterwards, we enumerated the system's requirements,
provided a proof of concept prototype based on available open source components,
and advanced to the deployment phase.

The optimality and validity of our proposition was shown through the response to
two hypothetical scenarios and an economic assessment of its related costs.
As a future work, we would like to investigate how to integrate  automatic
(or semi-automatic) request evaluation via decision aid tools and legal expert
systems.

Finally, we were concerned by the technical aspects of the proposition, it may be
interesting to investigate in details the associated organizational efforts such
the law enforcement policy.
\bibliographystyle{plain}
\biboptions{authoryear}
\nocite{*}
\bibliography{ref}
\end{document}